\documentclass[aps,prl,twocolumn,groupedaddress]{revtex4}
\usepackage{graphicx}% Include figure files
\usepackage{dcolumn}% Align table columns on decimal point
\usepackage{bm}% bold math

\begin{document}

%\preprint{huwei/2005NLC}

\title{Short-range interactions between strongly nonlocal spatial solitons}


\author{Wei Hu}
\email[These authors contributed equally to this work. ]{}%
\affiliation{Laboratory of Photonic Information Technology, South China Normal University,
Guangzhou 510631, P. R. China} %
\affiliation{Laboratory of Light Transmission Optics, South China Normal University, Guangzhou
510631, P. R. China}
\author{Shigen Ouyang}
\email[These authors contributed equally to this work. ]{} %
\affiliation{Laboratory of Photonic Information Technology, South China Normal University,
Guangzhou 510631, P. R. China} %
\affiliation{Laboratory of Light Transmission Optics, South China Normal University, Guangzhou
510631, P. R. China}
\author{Pingbao Yang}
\affiliation{Laboratory of Photonic Information Technology, South China Normal University,
Guangzhou 510631, P. R. China} %
\affiliation{Laboratory of Light Transmission Optics, South China Normal University, Guangzhou
510631, P. R. China}
\author{Qi Guo}
\email[]{guoq@scnu.edu.cn} %
\affiliation{Laboratory of Photonic Information Technology, South China Normal University,
Guangzhou 510631, P. R. China} %
\affiliation{Laboratory of Light Transmission Optics, South China Normal University,  Guangzhou
510631, P. R. China}
%



\date{\today}

\begin{abstract}
%From the conservation of power and momentum for the nonlocal
%nonlinear schr\"{o}dinger equation, we predict
A novel phenomenon is discovered that the short-range interaction
between strongly nonlocal spatial solitons depends sinusoidally on
their phase difference. The two neighbouring solitons at close
proximate can be inter-trapped via the strong nonlocality, and
propagate together as a whole. The trajectory of the propagation is
a straight line with its slope controlled by the phase difference.
%The tilting angle is obvious only when the separation
% between the solitons is in the region of beam width.
The experimental results carried out in nematic liquid crystals
agree quantitatively with the prediction. Our study suggests that
the phenomenon to steer optical beams by controlling the phase
difference could be used in all-optical information processing.
\end{abstract}

 % Beam trapping, self-focusing and thermal blooming

\maketitle


%\section{introduction}
Solitons are a common phenomenon appearing in many physical fields,
while the interactions of the solitons have great potential for much
wild applications.\cite{Stegeman1999} The strongly nonlocal spatial
soliton, which is also known the accessible soliton\cite{Snyder1997}
and is the self-trapped optical beam from the balance between
diffraction and nonlinearity propagating in nonlocal nonlinear media
under the condition of the strong
nonlocality\cite{krolikowski-pre-00}, has some significative
properties and has attracted more attentions in the last
decade\cite{Snyder1997,Nature2004,PRL2003,PRL2004}. Several strongly
nonlocal, referred also as highly nonlocal in some literatures (for
example, Refs.~\onlinecite{Snyder1997,PRL2003,PRL2004}),
 media have been found
in experiments, such as the nematic liquid crystal
\cite{PRL2003,PRL2004,Nature2004,APL2000}, the lead
glass\cite{PRL2005}, and the thermal nonlinear liquid\cite{PRL2006},
and the nonlinear gas of ions\cite{Suter1993}.

Nonlocality of nonlinear response may drastically modify the
properties of the solitons, specially their interactions. In a
strongly nonlocal case, it has been shown theoretically and
experimentally that the attraction can occur between the bright
solitons with any phase
difference\cite{Snyder1997,OL2002,Rasmussen2005,APL2006}, the
coherent or incoherent solitons\cite{Pecci_PRE2003,ShenM2005}, or
the dark solitons\cite{PRL2006,Nikolov2004}. In the contrary,
however, the interaction can be attractive only for two in-phase
local bright solitons\cite{Stegeman1999}. Both of the long-range
interaction\cite{Rotschild-NP-2006} and the short-range interaction
between the solitons can happen in strongly nonlocal nonlinearity,
%with the interaction distance much larger than the beam width,
while only the short-range interaction can occur in local
nonlinearity because the force between the local solitons decreases
exponentially with the separation between them\cite{Gordon-OL-2006}.

The fact well-known so far was
that\cite{Snyder1997,Rotschild-NP-2006,OL2002,Rasmussen2005,APL2006}
the interactions between the strongly nonlocal bright solitons are
independent of their phase difference. In this letter, we
%divide the interactions into two categories: a short-range interaction and a
%long-range interaction, and
differentiate the patterns of the short-range interaction and the
long-range interaction between the strongly nonlocal bright
solitons.
%, and clarify that the independence of the phase difference
%is only correct for the long-range interaction.
We show theoretically and experimentally that the short-range
interaction of the two strongly nonlocal bright solitons is
sinusoidally dependent on their phase difference.
%The definition of
%the short-range interaction and the long-range interaction will be
%given in the late discussion.
%In Kerr-medium, it means the two solitons escape in different angles.

%\section{theory part}

Let us consider a (1+2)-D model of an optical field that polarizes
linearly with an envelope $A$ and propagates in $z$-direction in the
medium with nonlocal nonlinearity:
\begin{equation}\label{GNNLSE}
2i k\partial_z A +\nabla_\perp^2 A + 2 k^2 \frac{\Delta n}{n_0} A  = 0,
\end{equation}
where $\nabla_\perp^2=\partial_x^2+\partial_y^2$, $k$ and $n_0$ are
the wave-vector and linear refractive index of the medium. The
nonlinear perturbation of refraction index $\Delta n(x,y,z)$ can be
generally expressed as
\begin{equation}\label{DELTA-N}
\Delta n=n_2 \int_{-\infty}^{\infty} R(x-x',y-y')|A(x',y',z)|^2
dx'dy',
\end{equation}
%
%We assume the response function $R(x-x',y-y')$ is shift-invariant, while the medium is large enough
%comparing to the width of soliton and the influence of the boundary condition is negligible weak.
where $R(x,y)$ is the real nonlinear response function of the medium. The normalized
condition, $\int_{-\infty}^{\infty} R(x,y) dxdy=1$, is chosen physically to make the
nonlinear index $n_2$ have the same dimensions as that in a local Kerr-medium. If
$R(x,y)$ is a delta function, $\Delta n=n_2 |A|^2$ and Eq.(\ref{GNNLSE}) becomes the
well-known nonlinear schr\"{o}dinger equations (NLSE) for the local Kerr-medium. The
Eq.(\ref{GNNLSE}) and Eq.(\ref{DELTA-N}), so-called the nonlocal nonlinear
schr\"{o}dinger equation (NNLSE), can model the beam propagation in most of nonlocally
nonlinear media discovered in experiments so far.

For the NNLSE, several well-known invariant integrals are important
for the theoretical analysis\cite{Rasmussen2005,Yakimenko2006}. The
first one is the power integral,
$P=\int_{-\infty}^{\infty}|A(x,y)|^2dxdy$, which results from the
energy conservation of an optical beam during the propagation in a
lossless medium. The second is the
momentum\cite{Rasmussen2005,Yakimenko2006}
\begin{equation}\label{M-DEFIN}
\vec{M} = \frac{1}{k}\int_{-\infty}^{\infty}A^*(-i \nabla_\perp) A {\rm d}x {\rm d} y.
\end{equation}
The momentum governs the movement of the mass center of the two
optical beams, i.e.
\begin{equation}\label{mass-center}
\frac{\partial \vec{r}_c(z)}{\partial z}= \frac{\vec{M}}{P},
\end{equation}
where the mass center $\vec{r}_c$ is $\vec{r}_c(z)=(1/P)\int
\vec{r}|A(x,y,z)|^2 dxdy$, and $\vec{r}=x\hat{e}_x+y\hat{e}_y$.
%If we choose $x_c(0)=0$,
The trajectory of mass center is a straight line with its slope
determined by Eq.~(\ref{mass-center}).

%It is same that $\tan\beta_y=M_y$.

%
\begin{figure}
\includegraphics[width=7cm]{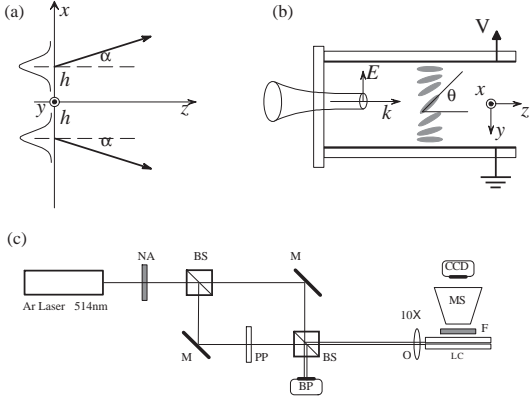}
\caption{The sketchs of (a) the two injective solitons, (b) the liquid crystal cell, and
(c) the experimental setup. NA, neutral attenuator; BS, beam splitters; M, plate mirror;
PP, parallel-face plate for adjusting the phase difference; O, 10$\times$ microscope
objective; LC, liquid crystal cell; MS, microscope; F, laser-line filter; BP, beam
profiler. \label{SKETCH}}
\end{figure}

Suppose the two simultaneously incident Gaussian solitons are
coplanar in $x-z$ plane, with a width $w_0$, a phase difference
$\gamma$ and a separation $d$($=2h$), as shown in
Fig.~\ref{SKETCH}(a), i.e.
\begin{eqnarray}\label{DGAUSS}
A(x,y,0)&=& A_0 \exp\left[-\frac{(x+h)^2+y^2}{2w_0^2}+ i k(x+h) \tan \alpha\right] \nonumber\\
 &+& A_0 e^{ i
\gamma} \exp\left[-\frac{(x-h)^2+y^2}{2w_0^2}- i k(x-h) \tan \alpha\right],
\end{eqnarray}
where the amplitude $A_0$ is large enough to make the two beams
propagate in soliton states\cite{Snyder1997,OL2002,PRL2003,PRL2004}.
For the input condition (\ref{DGAUSS}), the slope can be obtained
\begin{equation}\label{Q-DEFIN}
\frac{\tan\beta_x}{\Theta} = \frac{
(h/w_0)\exp\left[-\left(h/w_0\right)^2 -
\left(\tan\alpha/\Theta\right)^2\right]\sin\gamma}
{1+\exp\left[-\left(h/w_0\right)^2 -
\left(\tan\alpha/\Theta\right)^2\right] \cos\gamma},
\end{equation}and $\tan\beta_y=0$,
where  $\tan\beta_x$ and $\tan\beta_y$ are the slops in $x-z$ and
$y-z$ planes respectively,
%$M_x$ is the $x$-components of $\vec{M}$.where
and $\Theta=1/kw_0$ is the far-fields divergence angle of a Gaussian
beam.

\begin{figure}
\includegraphics[width=7cm]{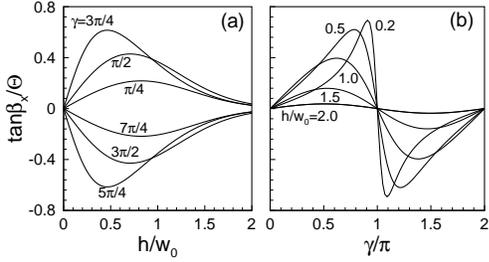}
\caption{The dependence of the slope on the distance $h$ (a) and phase difference
$\gamma$ (b) for two parallel-injected solitons.\label{ANGLE}}
\end{figure}

The slope of the line for the trajectory of mass center is greatly
dependent on the separation $d$ and the phase difference $\gamma$,
as shown in Fig.~\ref{ANGLE}, where we take $\alpha=0$, that is, two
solitons are parallel injected into the medium. Figure~\ref{ANGLE}
shows that $\tan\beta_x=0$ only when $\gamma=0$ or $\pi$ for
$h/w_0\leq2$, and $\tan\beta_x$ goes toward to zero when
$h/w_0\geq2$. $\tan\beta_x$ has significant value when $h$ is about
or smaller than the beam width $w_0$. In other words, when the
soliton separation $d$ is approximate or larger than four times the
soliton width $w_0$, the optical fields of the two solitons
%in Eq.(\ref{DGAUSS})
do not overlap so that $\tan\beta_x$ decreases to
zero, otherwise the two solitons are of an effective overlap and
$\tan\beta_x$ has a non-zero value changeable by the phase
difference.

It is important to emphasize that the above analytical result about
the movement of the mass center [Eq.~(\ref{Q-DEFIN})] is universal,
independent of the form of nonlinear response function $R$. This
means that no matter what the material is and how the degree of
nonlocality is,
%and how the input power of the beams is,
the movement of the mass center is the same for the initial
condition (\ref{DGAUSS}).

\begin{figure}
\includegraphics[width=7cm]{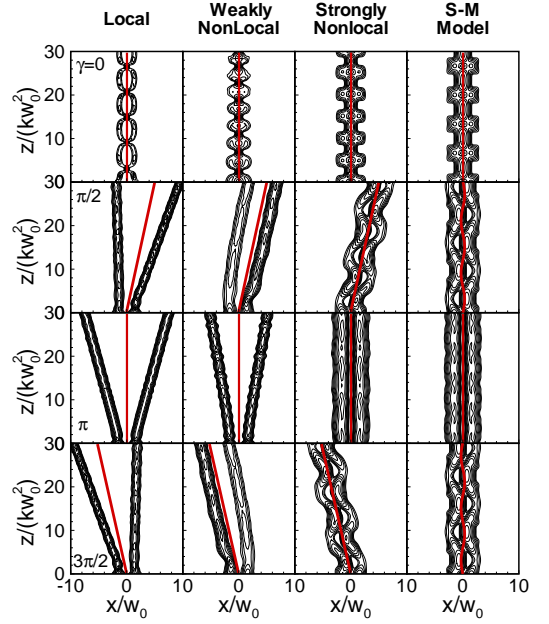}
\caption{The contour graph of the numerical propagation of
Eqs.~(\ref{GNNLSE}) and (\ref{DELTA-N}) for the two
parallel-injected solitons. The local case is shown in the first
(from the left to the right) column, and the two different nonlocal
cases with an exponential-decay response function given by
Eq.~(\ref{exponential-decay function}) are in the second and third
columns, respectively. The red solid lines show the movement of the
mass center of the two solitons. The results from the
Snyder-Mitchell model\cite{Snyder1997} with the same initial
condition are also given in the fourth column for comparison. The
phase difference between two solitons are $0$, $\pi/2$, $\pi$, and
$3\pi/2$, respectively (from the top to the bottom). \label{NUMNLC}}
\end{figure}

Although the momentum gives the movement of the mass center, it is
difficult to obtain the analytical solution of the beam propagation
for the initial condition (\ref{DGAUSS}). We carry out the numerical
simulation for local, weakly nonlocal, and strongly nonlocal
propagations, respectively, and only the (1+1)-D case of
Eqs.~(\ref{GNNLSE}) and (\ref{DELTA-N}) is simulated for the sake of
simplicity and also without the loss of generality. The (1+1)-D
model makes it possible to compare the propagations in the nonlocal
nonlinearity and in the local nonlinearity, and also provides an
exact enough description to the (1+2)-D coplanar propagation. The
results are shown in Fig.~\ref{NUMNLC}. The nematic liquid crystal
(NLC) is taken into consideration here for the consistency with our
experiment discussed later in this letter. The response function of
the NLC is the exponential-decay function
\begin{equation}\label{exponential-decay function}
R(x) =(1/2w_m)\exp(-|x|/w_w),
\end{equation}
for the (1+1)-D case\cite{Rasmussen2005}, and the zero-th order
modified Bessel function,
%
%\begin{equation}
$R(x,y) =(1/2\pi w_m^2)K_0(\sqrt{x^2+y^2}/w_m)$,
%\end{equation}
%
for the (1+2)-D cylindrical symmetrical case\cite{PRL2003,APL2006}, where $w_m$ is the
bias-voltage controllable characteristic length\cite{APL2006} of the response function
$R$ for the NLC. The ratio $w_m/w_0$ indicates the degree of
nonlocality\cite{krolikowski-pre-00}, which are chosen in simulation to be $0.47$ and
$10$ for weakly nonlocal and strongly nonlocal cases in Fig.~\ref{NUMNLC}, respectively.

%\newpage
Some interesting consequences can be obtained from
Fig.~\ref{NUMNLC}. For the local system described by the NLSE, the
two solitons attract each other only for the in-phase case
($\gamma=0$) and repel each other for other
occasions\cite{Stegeman1999}.
% , while the mass center of two
%solitons moves along the straight lines with the slope determined by
%the initial momentum $M_x$.
There also exists the power transfer between the solitons when
$\gamma$ is not equal 0 and $\pi$, as mentioned in
Ref.~\cite{Stegeman1999}. The force between the in-phase solitons is
always attractive, independent of the degree of nonlocality, as
shown in the first row; while the repelling force between the
solitons for the other phase difference cases becomes weaker, as the
degree of nonlocality increases. As a result, the two solitons with
an arbitrary phase difference can get attractive when nonlocality
becomes strong enough. For all of those propagations, however, the
movement of the mass center obeys the same regulation, a
straight-line trajectory with its slope given by
Eq.~(\ref{Q-DEFIN}). It is clear that the strong enough nonlocality
can make the two spatial solitons trap each other and propagate
together as a whole, going along the mass center trajectory, the
straight line, with the slope steered by their initial phase
difference. Interestingly, a similar phenomenon was mentioned
\cite{cohen-pre-2006} in the intermediate process\cite{explaining}
for dealing with incoherent solitons in ``fast'' nonlocal nonlinear
media.

It can be observed that the maximum value of the tilting angles
occurs when $\gamma$ approximates $\pi$ for small distance $h$. For
each $h$, two values $\gamma_{max}$ between $\pi/2$ and $3\pi/2$,
which is determined by $\cos\gamma_{max}= \pm\exp(-h^2/w_0^2)$,
makes $\beta_x$ reach the extrema (a maximum and a minimum,
respectively),
$\tan[(\beta_x)_{ext}]/\Theta=\pm(h/w_0)\exp(-h^2/w_0^2)
/\sqrt{1-\exp(-2h^2/w_0^2)}$. The maximum is for steering right and
the minimum for left. The smaller $h$, the larger tilt angle
$\beta_x$. The largest tilting angle is $\Theta/\sqrt{2}$. It means
the steering angle of whole beam are significance only for thin
beams.
%For a typical beam with $w_0=5\mu m$, the steering angle is
%only $\pm0.01\mathrm{rad}$.

Togetherness of Figs.~\ref{ANGLE} and \ref{NUMNLC} can result in one
more significant outcome. When $d<4w_0$, the two solitons have a
non-zero overlap and the slope $\tan\beta_x$ is no-zero also. In
this case, two solitons can be inter-trapped via the strong
nonlocality, and propagate together as a whole, going along the
diagonal line of the trajectory for their mass center. This is the
short-range interaction between the strongly nonlocal solitons,
which is phase-sensitive (controllable by their phase difference).
When $d>4w_0$, on the other hand, the two solitons never overlap and
the slope of the trajectory for their mass center tends toward zero.
In this case, two strongly nonlocal solitons undergo periodic
collisions in the coplanar
propagation\cite{Snyder1997,OL2002,Rotschild-NP-2006}, and spiral
about one another if they are initially skew to each
other\cite{Snyder1997,Rotschild-NP-2006}. Both processes have
nothing to do with their relative phase, as predicted first by
Snyder and Mitchell\cite{Snyder1997} and verified
experimentally\cite{OL2002,Rotschild-NP-2006}. This is the
long-range interaction between the strongly nonlocal solitons. As
the soliton separation $d$ increases from less to larger than
$4w_0$, the interaction will gradually transit from the short-range
pattern to the long-range one in the strongly nonlocal nonlinearity,
and vice versa. Only the short-range interaction, however, exists in
the local nonlinearity\cite{Gordon-OL-2006}.
%is a critical separation for this two kinds of
%interactions. and the long-range interaction between the strongly
%nonlocal solitons.

Two more points are given in the end of the theoretical part. It is
worth to note that the Snyder-Mitchell model\cite{Snyder1997} can
not give right prediction about the short-range interaction between
the strongly nonlocal solitons, as shown in the fourth column of
Fig.~\ref{NUMNLC}.
%The model is valid only for the in-phase and out-of-phase solitons. For
%other phase difference, they predict that two solitons attract each
%other and propagate along the $z$-axis, where the mass center
%oscillate harmonically around the $z$-axis as the solid red lines
%shown in Fig.~(\ref{NUMNLC}). They deal with the nonlinear change
%$\Delta n(x,y)$ but loss the influence of the detail distribution of
%light field $A(x',y')$ and responde function $(x-x',y-y')$.
Such a phase-controllable short-range interaction might have its
application in all-optical switching and routing.

\begin{figure}
\includegraphics[width=7cm]{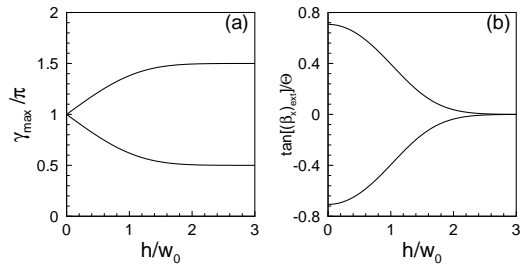}
\caption{(a) the phase difference $\gamma_{max}$ for the maximum slope angle vs the
distance $h$, and (b) the maximum of tilt angle $\tan[(\beta_x)_{ext}]$  vs the distance
 $h$. Both are for $\alpha=0$. \label{MaxANGLE}}
\end{figure}

%Another invariant integrals is the angular momentum,
%
%\begin{equation}
%L=\int_{-\infty}^{\infty} A^*[-i (x\frac{\partial}{\partial y} - y\frac{\partial}{\partial x})]A dx d y,
%\end{equation}
%which dominates the rotation of the optical field.  The conservation of the angular momentum has
%been used to explain the rotating of the optical soliton clusters in
%Kerr-medium\cite{Cluster-PRL}.  Here in Fig.(\ref{ANGULAR}), we show that the four soliton cluster
%with positive (top line) or negative(bottom line)  angular momentum rotates in anti-clock or clock directions.

%\begin{figure}
%\includegraphics[width=2.5cm]{left1.eps}
%\includegraphics[width=4cm]{left.eps}\\
%\includegraphics[width=2.5cm]{right1.eps}
%\includegraphics[width=4cm]{right.eps}
%\caption{The numerical simulation of the rotation of four solitons cluster in a (1+2)D nonlocal
%medium with a responding function.\label{ANGULAR}. }
%\end{figure}

%\section{In NLC}
In order to verify our prediction, we carried out the experiment in
the NLC.

%\subsection{Numerical of NLC}
The configuration of the NLC cell is the same as in the previous
works\cite{APL2006,APL2000,PRL2004}, as shown in
Fig.~\ref{SKETCH}(b). The optical field polarized in $y$-axis with
envelope $A$ propagates in $z$-direction. An external low-frequency
electric field $E_{RF}$ is applied in $y$-direction to control the
initial tilt angle of the NLC.
%The evolution of the paraxial beam $A$ and the tilt angle
%$\theta$ can be described by the equations in
%Ref.[\cite{peccianti-jnopm-o3,PRL2003}].

The experimental setup is illustrated in Fig.~\ref{SKETCH}(c). The laser beam from the
laser is split into two beams, then they are combined together with a small separation
through the other beam-splitter and launched into a $80\mu m$-thick NLC cell by a
10$\times$ microscope objective. The beam width at the focus $w_0$, the separation $d$,
and relative angle $2\alpha$ between the two beams are measured by an edged-scanning beam
profiler when the NLC cell is removed.  The cell is filled with the NLC TEB30A (from
SLICHEM China Ltd.), whose $n_\parallel = 1.6924$, $n_\perp = 1.5221$, $K\approx
10^{-11}N$, $\epsilon_a^{op}=0.5474$, and $\epsilon_a^{RF}=9.4$. The bias voltage on the
cell is set to $1.4$V, and then a pretilt angle is nearly $\pi/4$ in order to obtain
strong enough nonlocality and the lowest critical power of solitons\cite{APL2006}. The
launched power for each beam is fixed to 6mW, and two spatial solitons are obtained for
such high enough excitations. The parameters for the beams inside the NLC are calculated
from the measurement without NLC cell, i.e.
$w_0=2.2\mu$m, $d=2.25\mu$m%($h/w_0=0.51$)
, $\tan(2\alpha)=0.0076$, and the divergence angle $\Theta=0.0231$.

The phase difference between the two beams (solitons) is adjusted by
the rotation of a $1.8mm$-thick parallel-face plate, and measured
through the interference pattern by the beam profiler located on the
other branch after the second beam-splitter. First we find the
position of the plate while the phase difference is adjusted to $0$
(in phase), then we rotate the plate in small steps to increase the
phase difference $\gamma$.

\begin{figure}
\includegraphics[width=5cm]{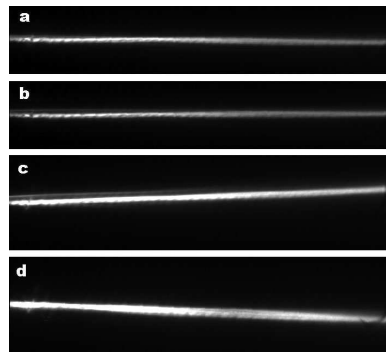}
\caption{Photos of the beam trajectories for the single soliton [(a)
and (b)] and the two solitons injected together [(c) and (d)]
propagating in the NLC cell. The phase difference between the two
solitons for (c) and (d) are about $\pi/2$ and $3\pi/2$,
respectively.}\label{PHOTO}
\end{figure}

We record the soliton trajectories for the different situations by
the CCD camera, as shown in Fig.\ref{PHOTO}. In Fig.~\ref{PHOTO} (a)
and (b), each of two solitons is alone launched into the NLC
respectively, and their trajectories are straight and horizontal.
When two solitons are injected simultaneously into the NLC, they
will propagate as a whole, and tilt (c) up or (d) down (actually in
$x$-direction). Since the separation is so small that two solitons
cannot be distinguished  by the microscope in our experiment, we
will see a whole beam, as a bound state, steered by the phase
difference $\gamma$.

\begin{figure}
\includegraphics[width=5cm]{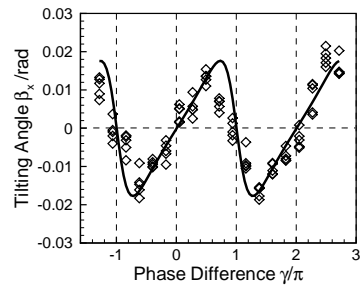}
\caption{The tilting angle of two beams vs. the phase difference
between them. Square points: experiment results, solid curve: the
theoretical fitting curve from Eq.~(\ref{Q-DEFIN}).\label{TILT}}
\end{figure}

In order to compare quantitatively our experimental observation with
our theoretical prediction, we give the variation of the tilting
angle with the phase difference $\gamma$ in Fig.~\ref{TILT}. For
each $\gamma$, we take five photos of the beam tilting angles to
minimize the jitter of the tilting angle resulting from that of the
laser source and the phase difference.
%The tilting angle of each photo are measured to be shown in
%Fig.\ref{TILT} as square points, and the theoretical predication is
%shown in solid line.
We can see that the experiment points locate
around the theoretical prediction with a relative small random
error. The error may mainly come from the slight jitter of the phase
difference $\gamma$. Except those random error, we can say that the
experiment results consist with the theoretical prediction very
well. The maximum tilt angle observed in experiment is about
$1.2^\circ$, approximate $0.6\Theta$.

%\section{conclusion and discussion}
In conclusion, we predict that the short-range interaction between
the strongly nonlocal solitons depends drastically on their phase
difference. The result is universal, independent of the different
nonlocal nonlinear media. The experiment carried out in the nematic
liquid crystal agrees quantitatively with the prediction. The
Snyder-Mitchell model can only give the right prediction for the
long-range interaction between the strongly nonlocal solitons.
%
%For the local nonlinearity, there exists only the short-range
%interaction between the solitons, but for

Unlike their local counterpart, the strongly nonlocal solitons can
exhibit both the short-range interaction and the long-range
interaction. The two kinds of interactions, however, have different
patterns, the former is phase-sensitive and the latter is not.
Therefore, each of the two phenomena would provide a means of
controlling light with light, and be thus potentially useful in
developing all-optical signal processing devices. They are, of
course, supposed to be applied in different situations.

This work was supported by the National Natural Science Foundation of China (Grant
No.10674050), and partially supported by the National Natural Science Foundation of China
(Grant Nos. 10474023 and 60278013). The authors would like to thank Prof. Li Xuan for her
supply of the NLC cell samples.

%These authors contributed equally to this work


\begin{thebibliography}{Reference}
%%Do not include separate BibTeX files; if BibTeX is used, paste the output here.
\bibitem{Stegeman1999}
%About the dependence on the phase-difference and the coherence, see,
G. I. Stegeman and M.
Segev,
%``Optical spatial solitons and their interactions: university and diversity'',
Science {\bf 286}, 1518 (1999) and references therein.

%\bibitem{TSKu2005} T.S. Ku, M.-F. Shih, A. A. Sukhorukov, and Y. S. Kivshar, \prl {\bf 94},
%063904 (2005).

%\bibitem{Anderson-pre-85}D. Anderson and M. Lisak, \pra {\bf32}, 2270 (1985).

\bibitem{Snyder1997}A. W. Snyder and D. J. Mitchell,
%``Accessible solitons,''
Science {\bf 276}, 1538 (1997).

\bibitem{krolikowski-pre-00}W. Krolikowski and O. Bang,
%``Solitons in nonlocal nonlinear media: exact solutions,''
Phys. Rev. E {\bf63} 016610 (2001); Q. Guo, B. Luo, and S. Chi,
%Optical beams in sub-strongly nonlocal nonlinear media: a
%variational solution,
Opt. Commun. {\bf 259}, 336 (2006); Q. Guo, B. Luo, F. Yi, S. Chi,
and Y. Xie, \pre {\bf69}, 016602 (2004).

\bibitem{PRL2003}C. Conti, M. Peccianti, and G. Assanto,
%route to nonlocality and obersvation of accessible soliton
\prl {\bf 91}, 073901 (2003).

\bibitem{PRL2004}C. Conti, M. Peccianti, and G. Assanto,
%" Observation of optical spatial solitons in highly nonlocal medium
\prl {\bf 94}, 113902 (2004).

\bibitem{Nature2004}M. Peccianti, C. Conti, G. Assanto, A. De Luca, and C. Umeton, Nature, {\bf
432}, 733 (2004).

\bibitem{APL2000}M. Peccianti, A. De Rossi, G. Assanto, A. De Luca, C. Umeton, and I. C. Khoo,
\apl {\bf 77}, 7 (2000).

\bibitem{PRL2005}C. Rotschild, O. Cohen, O. Manela, M. Segev, T.
Carmon,
%"solitons in nonlinear media with an infinite range of nonlocality:
% First observation of conherent elliptic solitons and vortex-ring solitons
\prl {\bf 95}, 213904 (2005)

\bibitem{PRL2006}A. Dreischuh, D. Neshev, D. E. Peterson, O. Bang, W. Kr$\acute{o}$likowski
%"Observation of attraction between dark soliton
\prl {\bf 96}, 043901(2006)


\bibitem{Suter1993} D. Suter, and T. Blasberg,
%"Stabilization of transverse solitary waves by a nonlocal response of nonlinear medium
\pra {\bf 45}, 4583 (1993).

\bibitem{OL2002}M. Peccianti, K. Brzdakiewicz, and G.
Assanto,
%"Nonlocal spatial soliton interactions in nematic liquid crystals",
\ol 27, 1460 (2002).

\bibitem{Rasmussen2005}P. D. Rasmussen, O. Bang, and Wieslaw Kr$\acute{o}$likowski,
%"theory of nonlocal soliton interaction in nematic liquid crstals"
\pre {\bf 72}, 066611 (2005).

\bibitem{APL2006}W. Hu, T. Zhang, Q. Guo, L. Xuan, S. Lan,
%" nonloclaity-controlled interaction of spatial solitons in nematic liquid crystals"
\apl {\bf 89}, 071111 (2006).

\bibitem{Pecci_PRE2003}M. Peccianti,and G. Assanto, \pre {\bf 65}, 035603(R) (2003).

\bibitem{ShenM2005} M. Shen, Q. Wang, J. Shi, Y. Chen, X. Wang,
%"nonlocal inconherent whight-light solitons in logarithmically nonlinear media
\pre {\bf 72}, 026604(2005).


\bibitem{Nikolov2004}N. I. Nekolov,
%"Attraction of dark soliton
\ol {\bf 29}, 286(2004).

\bibitem{Rotschild-NP-2006} C. Rotschild, B. Alfassi, O. Cohen and M. Segev, Nature Physics,
{\bf2}, 769 (2006).

\bibitem{Gordon-OL-2006}J. P. Gordon, \ol {\bf 8}, 596 (1983)
% Interaction forces among solitons in optical fibers J. P. Gordon Bell Laboratories,
%Holmdel, New Jersey 07733


\bibitem{Yakimenko2006}A. I. Yakimenko, V. M. Lashkin, and O. O. Prikhodko,
%"Attraction of dark soliton
\pre {\bf 73}, 066605 (2006).

%\bibitem{guo-pre-04}Q. Guo, B. Luo, F. Yi, S. Chi, and Y. Xie,
%``Large phase shift of nonlocal optical spatial solitons,''
%Phys. Rev. E {\bf69}, 016602 (2004).
%; N. Cao and Q. Guo, Acta Phys. Sin. {\bf54}, 3688 (2005).

%\bibitem{xie-oqe-05}Y. Xie and Q. Guo, Opt. Quant. Electron. {\bf36}, 1335 (2004).

%\bibitem{shen-science-97}Y. R. Shen,
%``Solitons made simple,'' Science {\bf 276} 1520 (1997).

%\bibitem{Segev1992} M. Segev, B. Crosignani, A. Yariv, B. Fischer,
%Spatial soliton in photorefracitve media
%\prl {\bf 68}, 923(1992)

%\bibitem{Cluster-PRL} A. S. Desyatnikov and Y. S. Kivshar, \prl {\bf 88}, 053901.

%\bibitem{Fratalocchi-mclc-2004}A. Fratalocchi, M. Peccianti, C. Conti, and G. Assanto, Mol.
%Cryst. Liq. Cryst. {\bf421}, 197 (2004).

%\bibitem{OL2005}M. Peccianti, C. Conti, and G. Assanto, \ol {\bf 30}, 415 (2005).


%\bibitem{Hutsebaut2004}X. Hutsebaut, C. Cambournac, M. Haelterman, A. Adamski, K. Neyts,
%" Single-component higher-order mode solitons in liquid crystls
%\oc {\bf 233}, 211(2004).



%\bibitem{peccianti-jnopm-o3}M. Peccianti, C. Conti, G. Assanto, A. De Luca, and C. Umeton,
%J. Nonl. Opt. Phys. Mat. {\bf12}, 525 (2003).

\bibitem{cohen-pre-2006}O. Cohen, H. Buljan, T. Schwartz, J. W. Fleischer, and M. Segev, \pre {\bf73}, 015601(R)
(2006).

\bibitem{explaining}In the Ref.~\onlinecite{cohen-pre-2006}, the propagation of incoherent
solitons in ``fast'' nonlocal nonlinear media was studied in two
steps: (1) analyzing the propagation within a very short time
interval (a time frame much shorter than the characteristic
fluctuation time) during which the beam can be treated as a coherent
speckled wave, and (2) calculating the propagation of the
time-averaged envelope. In the intermediate process (step 1), a
similiar phenomenon to steer self-trapped propagation by control
their phase difference is found to be exist for two overlap beams.
After step 2, however, such phase sensitivity is lost for the time
averaged envelope that is a really observed result.




%\bibitem{ICKhooBook}I.C. Khoo, Chap .6.

\end{thebibliography}
\end{document}